\documentclass[10pt, conference, letterpaper]{IEEEtran}
\IEEEoverridecommandlockouts
\usepackage{cite}
\usepackage{amsmath,amssymb,amsfonts}
\usepackage{url}
\usepackage{adjustbox,xspace}
\usepackage{subfigure}
\DeclareMathOperator*{\argmax}{arg\,max}

\usepackage{algorithm}
\usepackage{algorithmic}
\usepackage{graphicx}
\usepackage{textcomp}
\usepackage{xcolor}
\def\BibTeX{{\rm B\kern-.05em{\sc i\kern-.025em b}\kern-.08em
    T\kern-.1667em\lower.7ex\hbox{E}\kern-.125emX}}

\newcommand\systemname{mmHawkeye}

\newcommand{\review}[1]{{\color{black}{#1}}}
\newcommand{\reviewII}[1]{{\color{black}{#1}}}

\begin{document}
\title{mmHawkeye: Passive UAV Detection with a COTS mmWave Radar  
}
\author{
Jia Zhang$^1$, Xin Na$^1$, Rui Xi$^2$, Yimiao Sun$^1$, Yuan He$^{1}$ \\
$^1$School of Software \& BNRist, Tsinghua University, China \\
$^2$University of Electronic Science and Technology of China, China\\
$\{$j-zhang19, nx20$\}$@mails.tsinghua.edu.cn, ruix.ryan@gmail.com,\\
sym21@mails.tsinghua.edu.cn, heyuan@mail.tsinghua.edu.cn
}
\maketitle

\begin{abstract}

Small Unmanned Aerial Vehicles (UAVs) are becoming potential threats to security-sensitive areas and personal privacy. A UAV can shoot photos at height, but how to detect such an uninvited intruder is an open problem. This paper presents \systemname{}, a passive approach for UAV detection with a COTS millimeter wave (mmWave) radar. \systemname{} doesn't require prior knowledge of the type, motions, and flight trajectory of the UAV, while exploiting the signal feature induced by the UAV's periodic micro-motion (PMM) for long-range accurate detection. 
The design is therefore effective in dealing with low-SNR and uncertain reflected signals from the UAV. \systemname{} can further track the UAV's position with dynamic programming and particle filtering, and identify it with a Long Short-Term Memory (LSTM) based detector. We implement \systemname{} on a commercial mmWave radar and evaluate its performance under varied settings. The experimental results show that \systemname{} has a detection accuracy of 95.8\% and can realize detection at a range up to 80m.
\end{abstract}


\section{Introduction}

With the proliferation of small Unmanned Aerial Vehicles (UAVs), threats of UAVs arise, such as intrusion into personal space \cite{Harry}, illegal item delivery \cite{drug}, public safety threat \cite{liu2010long} and human injury \cite{Australian}, etc. Uninvited intrusion into personal space is most widely concerned, as it threats the privacy and safety of individuals. Without mandatory restrictions, a UAV can easily but illegally intrude into personal space at height to conduct activities such as candid photography or even theft. Since such UAVs are often very small and hard to spot with the naked eye, how to detect them becomes an extremely important and urgent problem.

A UAV detection system is desired to meet multiple goals: First, the detection approach should be passive, i.e., the detection process shouldn't require the cooperation of the UAV. Second, the system should be able to detect the presence of a UAV at height. Third, the detection system should be low-cost and easy to deploy, considering potentially a large population of ordinary users. Last but not least, the system should be generic to detect a variety of UAVs in various \review{illumination and noise conditions.}


Unfortunately, we find limitations of the existing approaches for UAV detection. Specifically, sound-based UAV detection \cite{acoustic_droneprint, he2023acoustic, yimiao2023aim, weiguo2023micnest} is easily interfered by complex environmental noise. The sound of UAVs attenuates fast in the air, so sound-based approaches generally have a limited detection range. Their performance further degrades when the UAV employs the noise reduction technique. Vision-based UAV detection \cite{video_vision} can work when the UAV is visible, but its accuracy and reliability are susceptible to illumination conditions and visual background. Thermal and \reviewII{Infrared Radiation (IR)} imaging cameras are possible options, but they are expensive and have limited coverage. There are also proposals of UAV detection based on RF signals \cite{RF_matthan, omnitrack}, which need to use special instruments to capture and analyze the communication of non-cooperative UAVs. Traditional radars \cite{ram, radar_lband} are too expensive and power-consuming. In short, none of the existing approaches is suitable for UAV detection in daily usage scenarios.


In this paper, we explore the feasibility of using a COTS (Commercial-Off-the-shelf) mmWave (millimeter Wave) radar for UAV detection. mmWave radar-based sensing \cite{rfwash, mmface, ambiear, rf-scg} has attracted a large body of research in the last few years. The UAV-reflected signals received by the radar contain rich information associated with the UAV. But it is a daunting task to accurately detect and identify a UAV from such signals, due to the following reasons: First, since a COTS radar has limited transmission power and the UAV is usually small and at height from the radar, the signals reflected from the UAV and received by the radar are very weak. Conventional approaches based on the signal intensity for target detection \cite{drone, jin2022passive, 2021detection, jin2023, 2022activity, he2019red, yang2022wiimg} become ineffective in such contexts. Second, the motions of a non-cooperative UAV (e.g., turning and hovering, etc.) are dynamic and unpredictable, making it extremely difficult to extract the UAV-reflected signals from the received signals. Third, the UAV-reflected signals contain both the inherent features of the UAV and the motion-related features, which are tightly coupled with one another. The above factors collectively lead to the very low signal-to-noise ratio (SNR) and the uncertainty of the UAV-reflected signals. 

To address the above challenges, we try to exploit a unique signal feature that can help us to extract and identify the reflected signals of the UAV. This feature is desired to be motion-independent, stable over time, consistent across different types of UAVs, and distinguishable from noise. Our finding is that the periodic micro-motion (PMM) of the UAV (such as propeller rotation, etc.) can be converted into stable and consistent features of the frequency of the reflected mmWave signals. Specifically, the periodic micro-motion always induces periodic frequency modulation of the reflected signal, resulting in a series of periodic peaks in the frequency spectrum. Based on this finding, we propose \systemname{}, a PMM-based UAV passive detection approach with a COTS mmWave radar. \systemname{} first extracts and enhances the periodic features with spectrum folding technology to enhance the signal SNR. A UAV tracking algorithm based on dynamic programming and particle filtering is designed to deal with unpredictable UAV motions. The extracted continuous features are then fed into an LSTM-based detector for accurate UAV identification.

Our contributions can be summarized as follows:

\textbullet\ To the best of our knowledge, \systemname{} is the first mmWave-based long-range UAV detection approach. By exploiting the PMM features, \systemname{} is able to detect and identify an uncooperative UAV with very weak signals. 

\textbullet\ We propose tailored algorithms based on the PMM feature to fully utilize the UAV's reflected signal, including feature extraction based on feature periodicity, UAV tracking with trajectory continuity and motion-independent UAV identification.

\textbullet\ We implement \systemname{} on the commercial radar (TI IWR6843ISKODS board) and conduct extensive experiments. The results demonstrate that \systemname{} achieves an average UAV detection accuracy of 95.8\% and an average relative range error of 0.9\% at a detection range up to 80m. 

The rest of the paper is organized as follows. Section  \uppercase\expandafter{\romannumeral2} presents the sensing model and the PMM feature, which are the theoretical foundation of this work. Section \uppercase\expandafter{\romannumeral3} elaborates on the design of \systemname{}. The implementation details and evaluation results are presented in Section \uppercase\expandafter{\romannumeral4}. 
Section \uppercase\expandafter{\romannumeral5} discusses practical issues and future work. Section \uppercase\expandafter{\romannumeral6} reviews the related work. Section \uppercase\expandafter{\romannumeral7} concludes this work. 

\section{Preliminaries}
\label{sec:model}
This section first introduces the sensing model and the PMM feature. Then we present observational results to demonstrate the feasibility of UAV detection with the PMM feature.


\subsection{The Sensing Model and the PMM Feature}
\begin{figure}[t]
     \centering
     \includegraphics[width=0.85\linewidth]{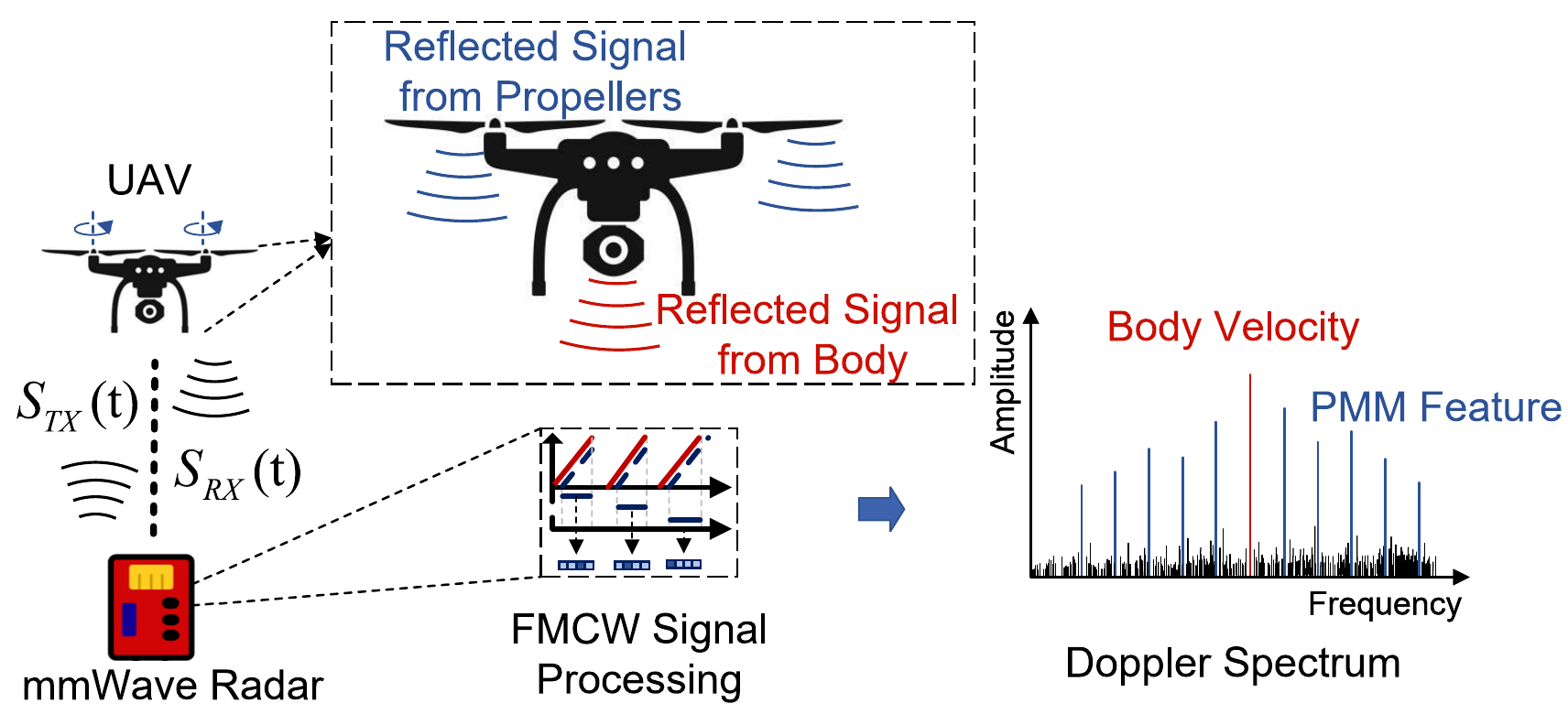}
     \vspace{-0.5em}
     \caption{The sensing model}
     \label{fig:model}
     \vspace{-0.5cm}
\end{figure}

As shown in Fig. \ref{fig:model}, the mmWave radar is deployed on the ground and faces upward. When a UAV is flying in the radar's sensing area, the signals emitted by the radar are reflected by the UAV and received by the radar. Specifically, a mmWave radar sends \textit{frequency-modulated continuous wave} (FMCW) chirp signals for range estimation and velocity measurement. If the impact of the UAV's micro-motion is ignored, the received signal includes only the reflected signal from the UAV's body. The distance between radar and UAV $R(t)$ can be obtained by calculating the \textit{beat frequency signal} $s(t)$, which is obtained by mixing the transmitted signal and the received signal:
\vspace{-0.1cm}
\begin{equation}
    s(t) = S_{Tx}^*(t)S_{Rx}(t) \approx \alpha exp[j4\pi(f_c + Kt)R(t)/c]    
\end{equation}
where $f_c$ and $K$ represent the starting frequency and the chirp slope of the FMCW signal, respectively. $\alpha$ is the propagation loss and $c$ is the speed of light. By conducting a \textit{Range-FFT} operation \cite{mmVib} on the samples of $s(t)$ during a chirp, the distance between the target and radar can be obtained. Then a \textit{Doppler-FFT} operation \cite{mmVib} is conducted on the Range-FFT results $S(t)$ in the corresponding bin across consecutive chirps to obtain the target's Doppler spectrum $S(f)$:
\vspace{0cm}
\begin{equation}
\begin{aligned}
    &S(t) = \alpha exp[j4\pi f_c R(t)/c] \\
    &S(f) = FFT(S(t)) = \alpha \delta(f - 2vf_c/c)
    \end{aligned}
\end{equation}
where $R(t) = R_0 + vt$, $R_0$ and $v$ are the current distance and the radial velocity, respectively. $\delta()$ is the Dirac delta function. The peak of the Doppler spectrum indicates the target velocity. With Range-Doppler-FFT, we can obtain the Range-Doppler spectrum and estimate the target's range and velocity.

When the micro-motion of the UAV is taken into account, the received signal includes the signals reflected from both the UAV's body and the propellers, as shown in Fig. \ref{fig:model}. Considering the small size of the UAV and the long distance between the UAV and the radar, the range difference between the propellers relative to the radar is negligible when the UAV is at height. Therefore we assume that all the propellers are in the same range bin. The number of propellers is denoted by $Q$. The blades of each propeller can be viewed as consisting of $P$ scattering points. Then the reflected signal from a UAV is the superposition of the signal reflected from the UAV's body and the signal reflected from each scattering point on the blades:
\vspace{-0.1cm}
\begin{equation}
\begin{aligned}
    S(t) = &\alpha exp[j4\pi f_c R(t)/c] + \\
           &\sum_{q=1}^{Q} \sum_{p=1}^{P} \beta_{pq} exp\{j4\pi f_c [R_{q}(t) + R_{pq}(t)]/c\}
\end{aligned}
\end{equation}
where $R_{q}(t) = R_q + vt$ is the distance between the $q$-th propeller rotor and the radar. $\beta_{pq}$ is the propagation loss. The distance between the $p$-th scattering point of the $q$-th propeller and the corresponding rotor projected to the radial direction $R_{pq}(t)$ can be calculated by:
\vspace{-0.1cm}
\begin{equation}
   R_{pq}(t) = r_{pq} * cos(\omega t + \phi_{pq}) * cos(\theta_{pq})
\end{equation}
where $r_{pq}$, $\omega$, $\phi_{pq}$ and $\theta_{pq}$ respectively denote the fixed distance between the $p$-th scattering point and the rotor, the rotational angular velocity of the propeller, the rotation initial phase of the $p$-th scattering point, and the angle between the propeller plane and the radial velocity direction.

If we perform a Doppler-FFT operation on this reflected signal, its Doppler spectrum can be represented as:
\vspace{-0.1cm}
\begin{equation}
\begin{aligned}
    S(f) = &\alpha \delta(f - 2vf_c/c) + \\
           &\sum_{q=1}^{Q} \sum_{p=1}^{P} \sum_{m=-\infty}^{+\infty} \gamma_{pqm} \delta(f - 2vf_c/c - \omega m/{2\pi})
\end{aligned}
\label{eq:doppler}
\end{equation}
where $\gamma_{pqm}$ denotes the $m$-th frequency loss of a scattering point, which is a complex function of $m$, $\beta_{pq}$, $r_{pq}$, $\phi_{pq}$ and $\theta_{pq}$. For more calculation details, the reader can refer to the related work \cite{analysis}. This formula shows that there will be a series of periodic peaks centered on the body velocity peak in the Doppler spectrum. We call these peaks \textbf{the PMM feature}:
\begin{equation}
\begin{cases}
&\text{Peak Pos}: 2vf_c/c + \omega m/{2\pi} \quad \text{$m \in Z$}\\
&\text{Peak value}: \sum_{q=1}^{Q} \sum_{p=1}^{P} \gamma_{pqm} \quad \text{$m \in Z$}
\end{cases}
\end{equation}
The interval between the peaks is the same ($\omega/{2\pi}$) and is determined by the propeller rotation velocity. The peak values are more complex and related to the UAV relative position ($\beta_{pq}$, $\theta_{pq}$) and the UAV structure ($r_{pq}$, $\phi_{pq}$, $P$, $Q$). 

With the derivation of the PMM feature, we can find that the PMM feature exists stably during the UAV's flight. The structure and position of a UAV only affect the peak values rather than the peak intervals. \review{This means that the number of propellers, the number of blades and the blade length don't affect the peak intervals.} Such a stable feature can be used to detect various UAVs. Without loss of generality, we use a six-wing UAV as an example in later sections.




\subsection{PMM in Reality}
We give the observations on PMM to demonstrate the feasibility of UAV detection with the PMM feature.

\begin{figure}[t]
    \centering
    \subfigure[The ascending UAV]{
    \includegraphics[width=0.9\linewidth]{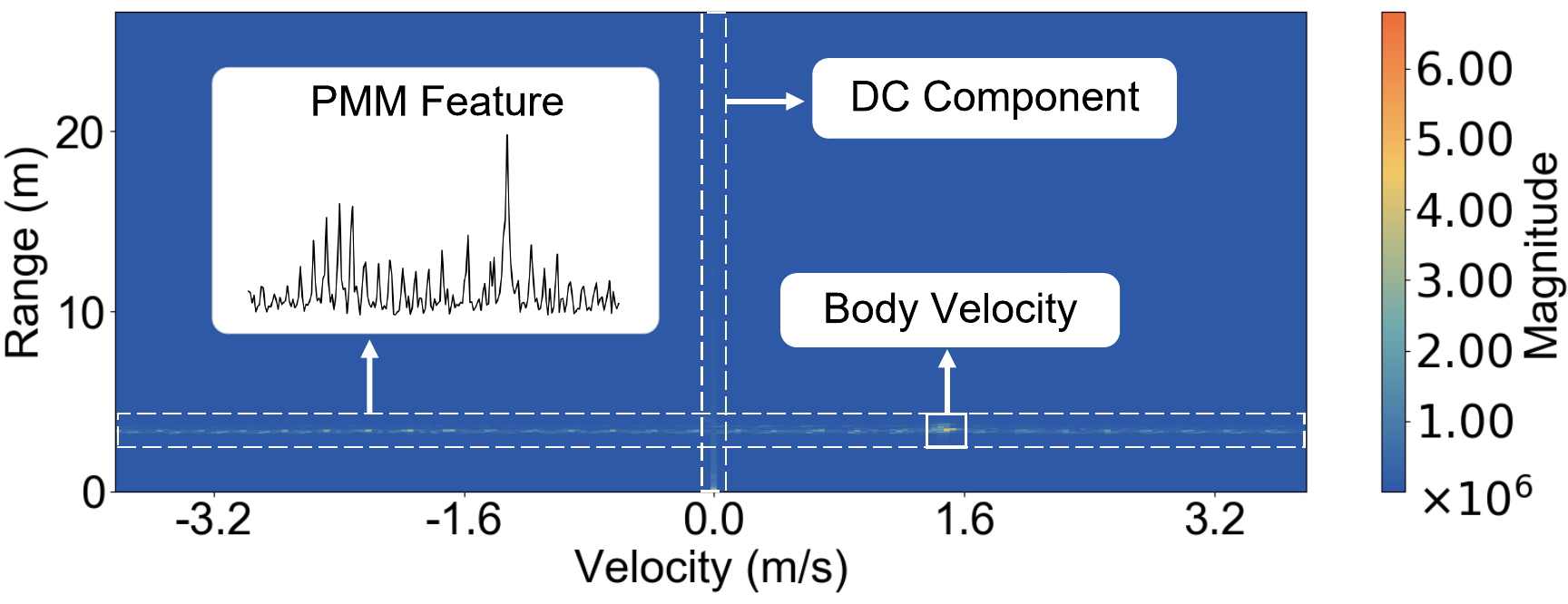}
    }
    \subfigure[The hovering UAV]{
    \includegraphics[width=0.9\linewidth]{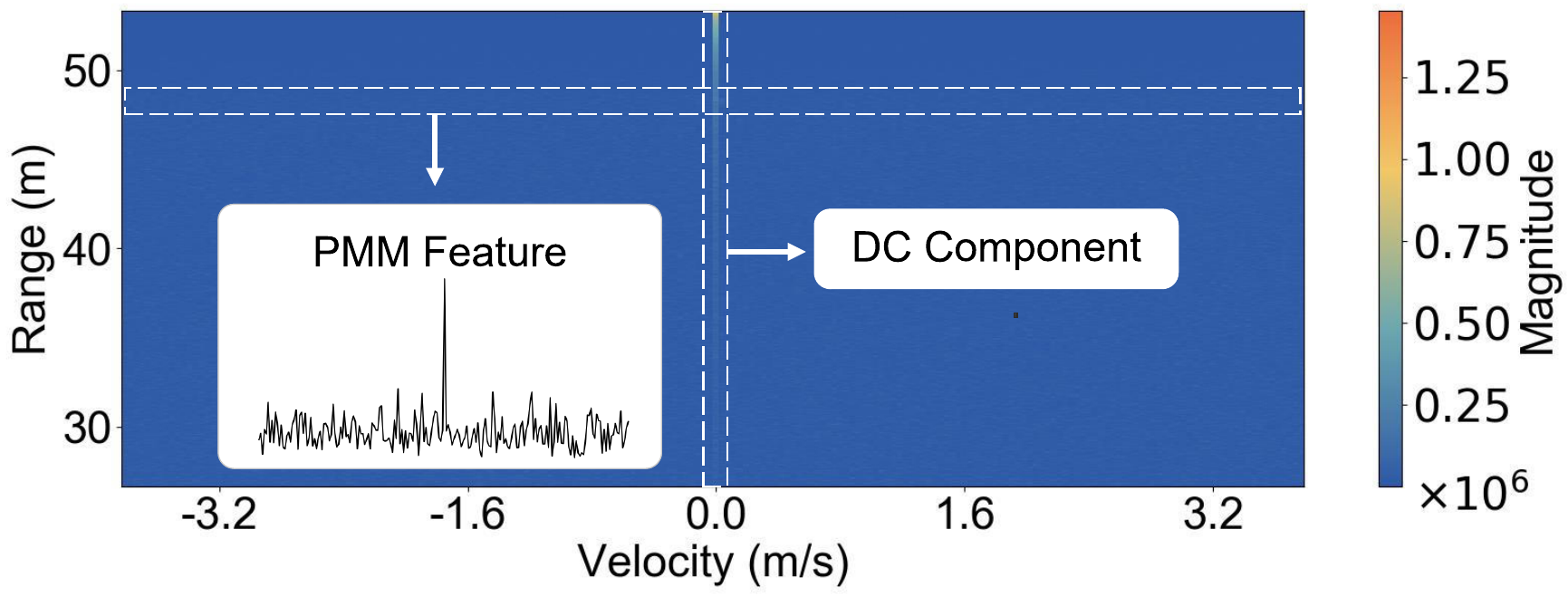}
    }
    \vspace{-0.2cm}
    \caption{The Range-Doppler spectrum of the UAV}
    \label{fig:preliminary}
    \vspace{-0.5cm}
\end{figure}

In our experiments, a commercial mmWave radar (TI IWR6843ISKODS board) \cite{6843} is deployed on the ground and faces upward. 
A six-wing UAV with three blades per propeller is steered above the mmWave radar. The UAV is first controlled to ascend at a constant speed of about 1.5m/s and then hover at a range of  48m.
We respectively select a segment of the reflected signal from these two stages and perform Range-Doppler-FFT on them. Fig. \ref{fig:preliminary}(a) and Fig. \ref{fig:preliminary}(b) show the Range-Doppler spectrum of the ascending UAV and that of the hovering UAV, respectively. 
The results show that: (1) Whether the UAV is moving or hovering, the PMM feature always stably appears in the range bin where the UAV is located. In contrast, the reflection intensity and the radial velocity features are less stable as they are more easily affected by the relative distance and the UAV's motions. (2) When the distance between UAV and radar is far, the PMM feature shrinks and distorts, due to the degraded quality of the reflected signal. Nevertheless, its periodicity still exists stably. Note that the highest peak of the PMM feature in Fig. \ref{fig:preliminary}(b) corresponds to the \reviewII{Direct Current (DC)} components. In this case, we try to exploit the periodicity of the UAV's PMM feature to extract and identify the reflected signal of the UAV.


\begin{figure*}[t]
     \centering
     \includegraphics[width=0.7\linewidth]{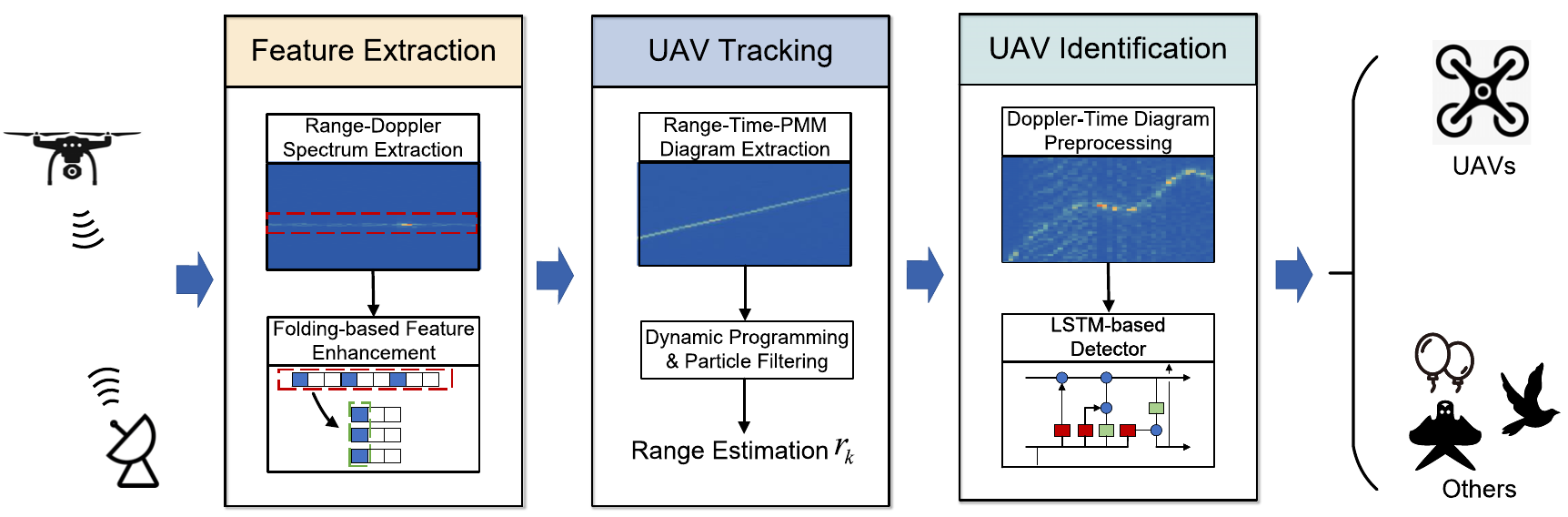}

     \vspace{-0.5em}
     \caption{The overflow of \systemname{}}
     \label{fig:overflow}
     \vspace{-0.5cm}
\end{figure*}
\section{\systemname{} Design}
This section starts with an overview of  \systemname{}, and then introduces three key modules of the design respectively.
\subsection{Overview}
\systemname{} solves the UAV detection problem in three steps, which correspond to the three key modules in its design, namely feature extraction, UAV tracking and UAV identification. Fig. \ref{fig:overflow} shows the overview of \systemname{}. 

\textbullet\ \textit{Feature extraction.} \systemname{} continuously extracts the Range-Doppler spectrums with Range-Doppler-FFT. To distinguish the PMM feature from the environment noise in each Range-Doppler spectrum, We perform spectrum folding on each Doppler spectrum. If the PMM feature exists, it will be significantly enhanced by spectrum folding and the folding result can be used for further tracking and detection.




\textbullet\  \textit{UAV tracking.} With feature extraction, \systemname{} continuously extracts the folding results in Range-Doppler spectrums, which collectively form the Range-Time-PMM diagram. To cope with the impact of unpredictable UAV motions, we design a tracking algorithm based on dynamic programming and particle filtering to obtain the tracking result from the Range-Time-PMM diagram.


\textbullet\ \textit{UAV identification.} After obtaining the target's tracking result, \systemname{} continuously extracts the Doppler spectrums from the target's location, forming the Doppler-Time diagram. After preprocessing the diagram via DC removal and feature alignment, the Doppler-Time segments are fed into an LSTM-based detector for UAV identification.



\subsection{Feature Extraction}

The feature extraction module employs the spectrum folding technique to amplify the difference between the PMM feature and the environment noise. Below is the detail of this module.

As we mentioned before, we can obtain the Range-Doppler spectrum by performing Range-Doppler-FFT on the reflected mmWave signals. Specifically, we use $\{D_1, D_2,...,D_R\}$ to represent the Doppler spectrums in a Range-Doppler spectrum. $D_i$ and $R$ represent the Doppler spectrum in the $i$-th range bin and the number of range bins, respectively. The number of Doppler bins in a Doppler spectrum is $L$. When the UAV appears in the $i$-th range bin, the PMM feature will appear in the corresponding Doppler spectrum $D_i$. However, considering the unpredictable propeller rotation velocity, the period of the peaks in the PMM feature is uncertain and variable. Those periodic peaks may even be buried in environment noise due to the low SNR of the reflected signal. 

\begin{figure}[t]
     \centering
     \includegraphics[width=0.8\linewidth]{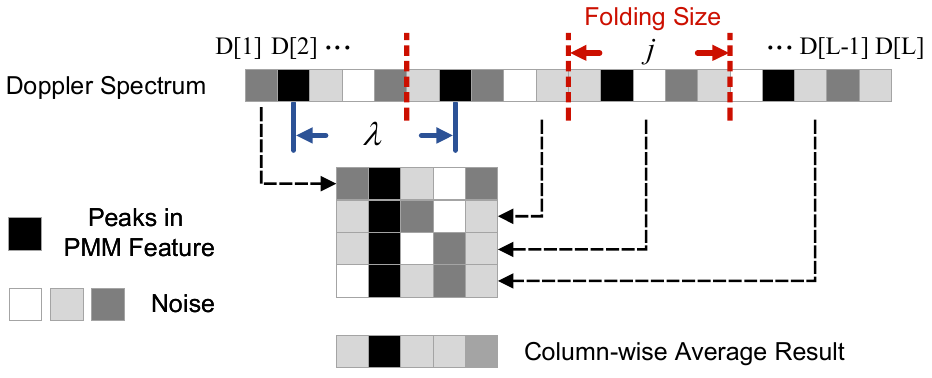}
              \vspace{-0.5em}
     \caption{The folding process of the PMM feature}
     \label{fig:folding}
         \vspace{-0.6cm}
\end{figure}

We employ the folding technique \cite{freebee, guo2020wizig} to extract and enhance the PMM feature, which is used to find signal periodicity under noise.  An example of the folding process is shown in Fig. \ref{fig:folding}, where the values of Doppler bins are represented by boxes. The black boxes represent the periodic peaks in the PMM feature and the other boxes represent the noise. The interval $\lambda$  between adjacent peaks in the Doppler spectrum is 5 and the number of Doppler bins $L$ is 20. If we exactly fold the Doppler spectrum into a matrix with $j=\lambda$ columns, the periodic peaks will align in a column and there will be a significantly enhanced peak in the column-wise averaged result. For clarity, the maximum value in the column-wise averaged result and the number of folding columns are referred to as the folding value and the folding size, respectively. When the folding size $j$ is not equal to $\lambda$, the folding value will decrease rapidly as the peaks are not aligned.

Note that the interval between adjacent peaks in the PMM feature is determined by the propeller rotation velocity, it is unpredictable and may change at any time. To find the right folding size, we traverse all integers in an empirical range $[j_{min}, j_{max}]$ and calculate the corresponding folding values, where $j_{min}$ and $j_{max}$ represent the minimum and maximum folding size, respectively. \review{They are set to $2$ and $20$ according to the propeller rotation velocity range and the Doppler spectrum resolution.} With the fixed traversing range, the computation time can be controlled. The folding value of Doppler spectrum $D_i$ with folding size $j$ can be calculated by:
\begin{equation}
F_{i}(j) = \max_{1 \leq k \leq j} \frac{\sum_{1 \leq m \leq M} {D_{i}[k + (m - 1) * j]}}{M}
\end{equation}
where $k$ and $m$ denote the folding column index and the row index in the folding matrix, respectively. The number of rows in the folding matrix is represented by $M$, which can be calculated by $M =\lfloor \frac{L}{j} \rfloor$.

The largest folding value is selected to be the folding result, which increases significantly when the PMM feature exists. The folding result of the Doppler spectrum $D_i$ can be calculated by:
\begin{equation}
    P_i = \max_{j_{min} \leq j \leq j_{max}} F_{i}(j)
\end{equation}

In this way, \systemname{} can calculate the folding result of each Doppler spectrum. Since the PMM feature of the UAV has a stable periodicity, its folding result is much larger than that of the random environment noise. These folding results can be further used for UAV tracking and detection. 

\subsection{UAV Tracking}
The UAV tracking module first preprocesses the folding results with spectral subtraction, so that the impact of the static background noise is mitigated. Then the UAV tracking is realized through dynamic programming and particle filtering, where the unpredictable UAV motions and the local dynamic noise are taken into account and appropriately dealt with.

To estimate the UAV trajectory, \systemname{} continuously extracts the folding results with feature extraction. These folding results form a Range-Time-PMM diagram, denoted by the R-PMM. Suppose that there are $T$ Range-Doppler spectrums and each Range-Doppler spectrum has $R$ range bins, 
the value of R-PMM(r, t) represents the folding result of the $r$-th range bin in the $t$-th Range-Doppler spectrum. As the Doppler spectrum of the UAV has a larger folding result at each moment, \review{the UAV trajectory corresponds to a series of range bins that contain larger values in the R-PMM. These range bins can form a maximum path in the R-PMM whose cumulative folding result over time is the maximum among all paths. Therefore, } we can track the UAV by searching for the maximum path in the R-PMM.

\review{Considering that the intensity of the Doppler spectrum is proportional to the reflected signal intensity, there are different static background noise intensities in different range bins in the R-PMM.} To mitigate the impact of such noise, we employ the spectral subtraction algorithm \cite{advanced} to preprocess the R-PMM. 
The main idea is to subtract the estimation of the average background noise spectrum from the noisy R-PMM. Specifically, the average background noise spectrum can be estimated by:
\vspace{-0.1cm}
\begin{equation}
    N(r) = \frac{1}{T} \sum_{t=1}^{T} N(r,t)
\end{equation}
where $N(r,t)$ is the R-PMM measured in the background noise. The gain of the background noise is calculated as a normalized projection of the noisy R-PMM onto the background noise spectrum:
\vspace{-0.1cm}
\begin{equation}
    G(t) = \sum_{r=1}^{R} \frac{N(r)S(r,t)}{||N||^2}
\end{equation}
where $S(r,t)$ is the measured R-PMM in the tracking phase and $||N|| = \sqrt{N(1)^2+N(2)^2+...+N(R)^2}$ is the Euclidean norm of the noise spectrum. Finally, the background noise can be removed from the measured R-PMM as follows:
\begin{equation}
    S^{'}(r,t) = S(r,t) - G(t)N(r) 
\end{equation}

In this way, the range-related static background noise can be removed and the preprocessed R-PMM can be used to find the maximum path, which corresponds to the UAV tracking result. \review{The maximum path $g^*$ can be obtained by solving:
\vspace{-0.1cm}
\begin{equation}
\label{problem}
\begin{aligned}
    &g^* = \argmax_{g}(\sum_{t=1}^{T}S^{'}(g(t), t)) \\
\end{aligned}
\end{equation}
where $g = (t, g(t))_{t=1}^T$ is denoted as a path.}


However, since the folding result characterizes the PMM feature of the UAV, it may change rapidly due to unpredictable UAV motions. On the other hand, impacted by the complex environment and imperfect hardware, there may still be local dynamic noise in the vicinity of the UAV trajectory in the R-PMM. Both of them lead to degradation in tracking accuracy.

Considering that the UAV trajectory always changes continuously, \systemname{} utilizes the trajectory continuity to reduce the tracking errors. Specifically, assume that the maximum flight speed of the UAV is $V_{max}$ and the duration of the Range-Doppler spectrum is $T_d$, the maximum range bin variation of the UAV in the adjacent columns of the R-PMM can be calculated by: 
\vspace{-0.1cm}
\begin{equation}
    K = \lceil \frac{V_{max}*T_d}{R_{res}} \rceil 
\end{equation}
where $R_{res}$ represents the range resolution of the mmWave radar. The path in the R-PMM corresponding to the UAV trajectory always satisfies this constraint. Therefore, the UAV tracking problem can be further transformed into the problem of finding a constrained maximum path in the R-PMM, where the variation of adjacent columns in the path does not exceed $K$. This constrained maximum path can be found by solving:

\begin{equation}
\label{problem}
\begin{aligned}
    &g^* = \argmax_{g}(\sum_{t=1}^{T}S^{'}(g(t), t)) \\
    &s.t. \; |g(t) - g(t - 1)| \leq K \\
\end{aligned}
\end{equation}

This problem can be solved by dynamic programming. Specifically, we first define the score at bin $(r, t)$ as the constrained maximum cumulative folding result, which can be calculated as:

\begin{equation}
    \theta(r, t) = \max_{k \in [-K, K]} \theta(r + k, t - 1) + S^{'}(r, t)
\end{equation}
Since $\theta(r, t)$ considers both the trajectory continuity and the previous cumulative folding results, its calculation process obtains the constrained optimal track through the bin $(r, t)$. To obtain the entire constrained maximum path, we first find the bin $(T, g^{*}(T))$ in the last column that contributes to the maximum score. Then the rest of the path can be obtained by:
\vspace{-0.1cm}
\begin{equation}
\begin{aligned}
    &g^{*}(t) = \argmax_{k \in [-K, K]} \theta(g^{*}(t + 1) + k, t) + g^{*}(t + 1) \\
    &\forall \; t = T - 1, T - 2,..., 1 \\
\end{aligned}
\end{equation}
This backtracking procedure provides the constrained maximum path $g^{*}$, which is the optimal solution for Eq. \ref{problem}. In this way, we can obtain the UAV tracking result. 




\systemname{} further applies the particle filter \cite{particle} to the tracking result to reduce the tracking error. The particle filter can estimate the target state by combining the observation and the prediction. 
Specifically, the state in our particle filter includes the range and the velocity of the UAV, and the observation is the tracking result. We initialize 5000 particles with uniform distribution and use the multinomial resampling algorithm as the particle resampling method.

\subsection{UAV Identification}

\begin{figure}[t]
     \centering
     \includegraphics[width=0.6\linewidth]{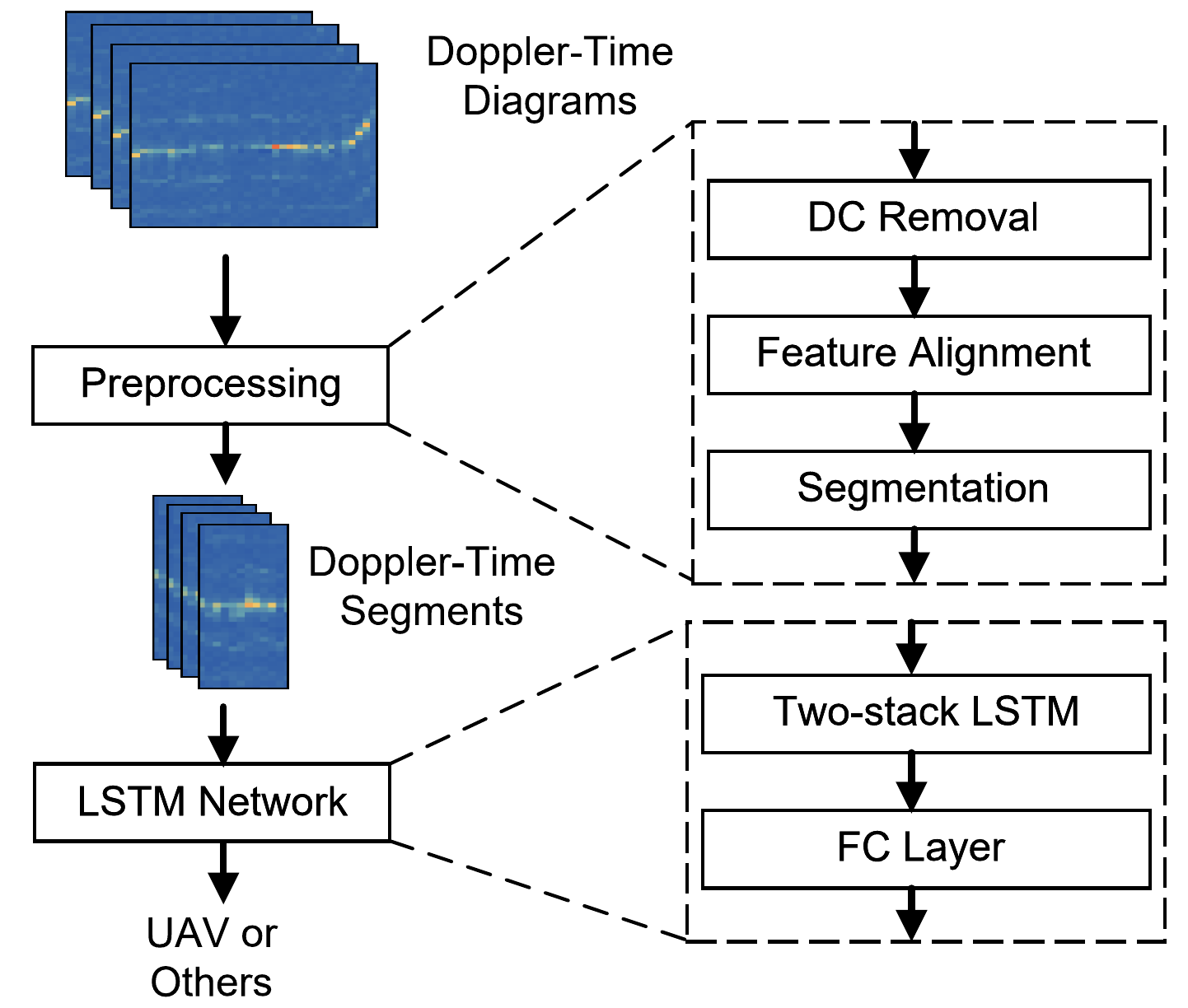}
              \vspace{-0.5em}
     \caption{The process of UAV identification}
     \label{fig:detection}
         \vspace{-0.5cm}
\end{figure}

Now we show how to utilize the PMM features to identify a UAV. We first extract the target's Doppler spectrums from a series of Range-Doppler spectrums according to the tracking result. Then the Doppler spectrums are fed into an 
LSTM-based) detector for UAV identification. The process of UAV identification is shown in Fig. \ref{fig:detection}.

With the tracking result and the Range-Doppler spectrums, we can extract a series of Doppler spectrums from where the target is located, forming a Doppler-Time diagram. When the target is a UAV, this diagram will contain unique and continuous PMM features which can be used to distinguish the UAV from other objects.

We first remove the DC noise from the Doppler-Time diagram. Considering that the body velocity of the UAV is close to zero when it hovers, the corresponding DC component contains the body velocity peak. To preserve the body velocity peak, we average the DC components of the Doppler spectrums where the body velocity peak is not close to the DC component. Then the average value is subtracted across all the DC components. Besides, the unpredictable body velocity determines the center of the PMM feature and prevents us from exploiting the periodicity of the PMM feature. Therefore We devise a feature alignment algorithm on the Doppler-Time diagram to align each PMM feature center to the Doppler spectrum center. When the body velocity peak of the Doppler spectrum is not in the DC bin, we shift the entire Doppler spectrum along the direction from the body velocity peak to the DC bin and complement it by linear interpolation. 


The preprocessed Doppler-Time diagram contains the PMM features. However, since the tracking result may have errors, especially when the UAV is at height. It is not always reliable to directly use the obtained Doppler-Time diagram. To reduce the impact of tracking errors, we first split the Doppler-Time diagram into fixed-length segments. Then we compare the maximum folding result of each segment with an empirical threshold. When the maximum folding result is less than the threshold, we consider that the segment does not contain PMM features and discard it. The threshold is set to 30000 according to our extensive empirical experiments.



\review{
Finally, the preprocessed Doppler-Time segments are used to identify the UAV from other objects. Considering that when there are objects such as birds or balloons in the vicinity of the radar, their Doppler-Time segments may have high folding results and uncertain folding sizes due to the environmental noise, the simple classification methods based on statistical features of PMM, such as peak interval, mean value, etc., may be easily disturbed by such segments and severely degrade. Therefore, we design an LSTM-based detector to solve such a binary classification problem.} The LSTM network \cite{lstm} is a classic recurrent neural network that is suitable to process the sequences of data and has excellent performance on recognition tasks. 
Considering that each time slot in the Doppler-Time diagram contains $L$ Doppler bins, the input dimension of our LSTM network is set to $L$. Our network contains two stacked LSTM layers and the hidden state size is set to 128. A fully connected layer is used to map the hidden state to the identification results, i.e., UAVs or other objects. We select the cross entropy loss function to train our network. \reviewII{Considering that the peak intervals in PMM features are not affected by UAV types, our network can detect various UAVs with small training data.} 




\section{Implementation and Evaluation}
In this section, we introduce the implementation of \systemname{} and evaluate the performance of our prototype under different settings.

\begin{figure}[t]
    \centering
    \includegraphics[width=0.6\linewidth]{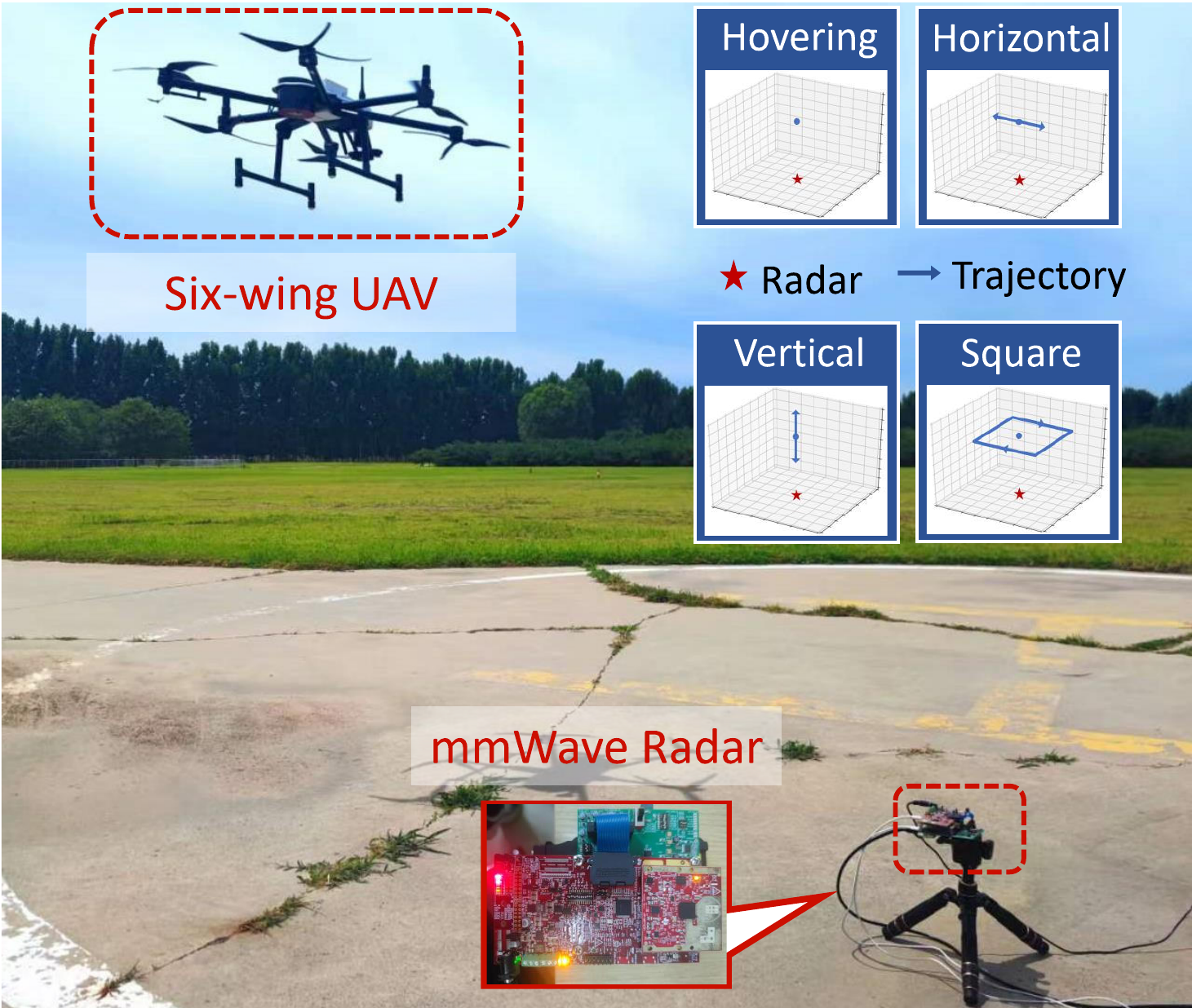}
    \caption{The experiment scenario}
    \label{fig:experiment}
    \vspace{-0.5cm}
\end{figure}

\begin{figure*}[ht]
  \centering 
   \begin{minipage}{0.22\textwidth}
 \centering
 \includegraphics[width=\textwidth]{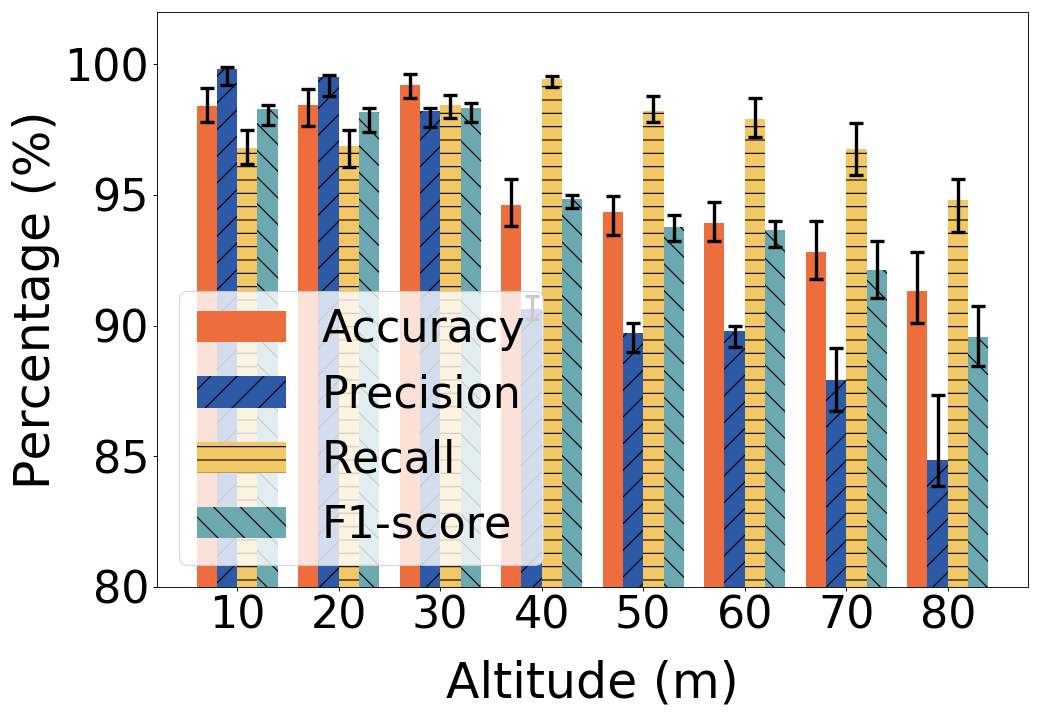}
 \vspace*{-0.5cm}
  \caption{\review{The overall performance of UAV detection}}
  \label{fig:overall_detection}
\end{minipage}
\hspace{0.2cm}
\begin{minipage}{0.22\textwidth}
\centering 
\includegraphics[width=\textwidth]{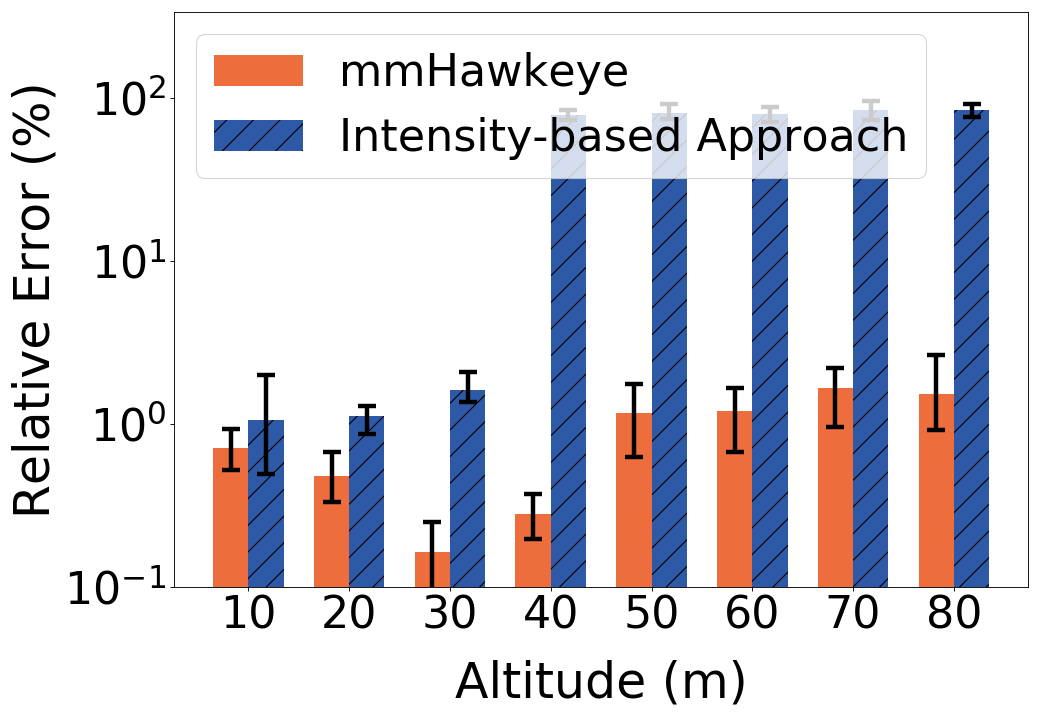}
\vspace*{-0.5cm}
\caption{The overall performance of UAV tracking}
  \label{fig:overall_range_error}
\end{minipage}
\hspace{0.2cm}
\begin{minipage}{0.22\textwidth}
\centering 
\includegraphics[width=\textwidth]{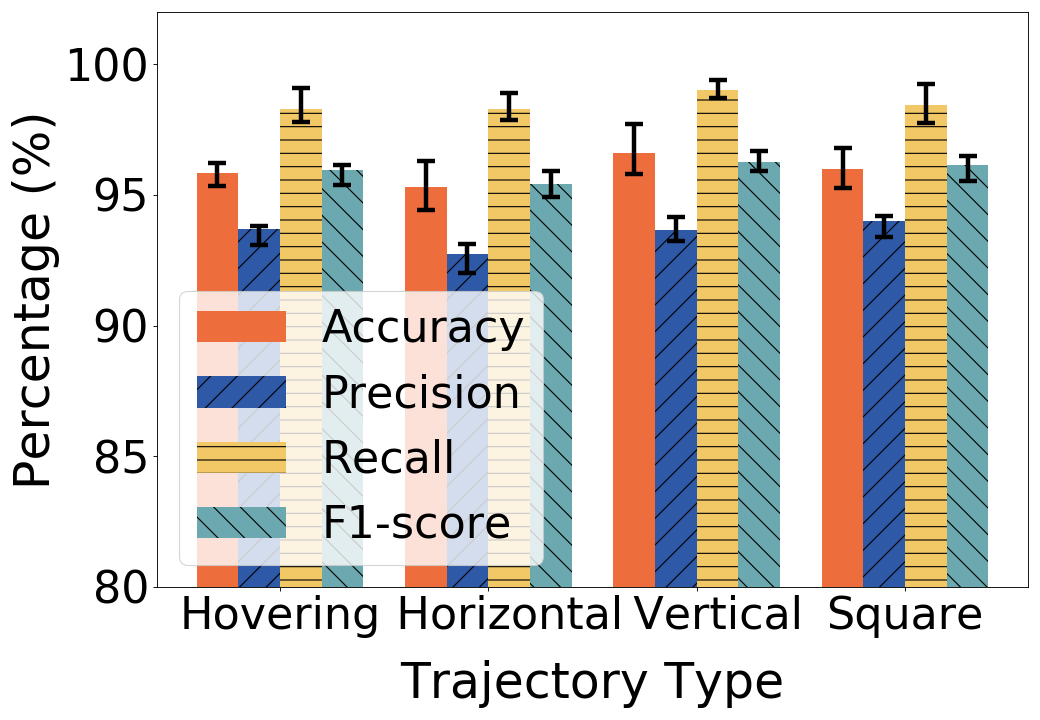}
\vspace*{-0.5cm}
\caption{\review{The impact of UAV trajectory on detection results}}
  \label{fig:trajectory_detect_error}
\end{minipage}
\hspace{0.2cm}
 \begin{minipage}{0.22\textwidth}
 \centering
 \includegraphics[width=\textwidth]{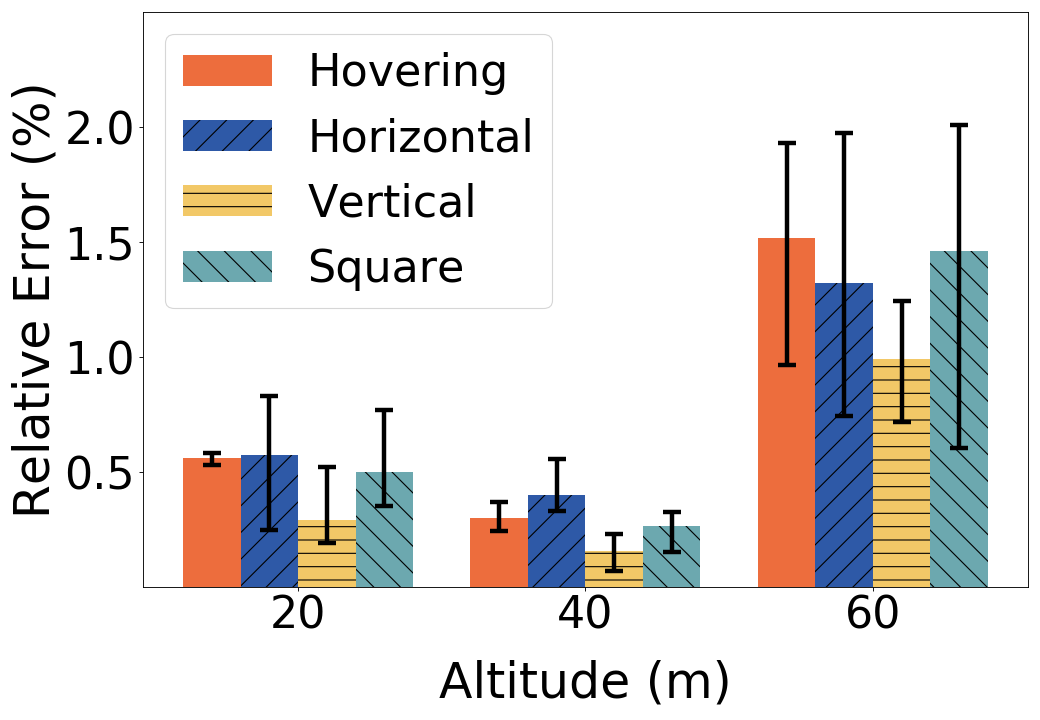}
 \vspace*{-0.5cm}
  \caption{The impact of UAV trajectory on tracking results}
  \label{fig:trajectory_range_error}
\end{minipage}
    \vspace{-0.5cm}
\end{figure*}

\subsection{Implementation}
\label{sec:implementation}
We implement \systemname{} on a commercial mmWave radar Texas Instruments IWR6843ISKODS \cite{6843}. There are 3 Tx antennas and 2*2 Rx antennas on the radar board. In our implementation, we let three TX antennas take turns transmitting FMCW signals starting at 60.25GHz with 1.92GHz bandwidth, and all Rx antennas receive the reflected signals. The duration of a single chirp is 900us and each frame includes 100 chirps. The frequency slope of the FMCW signal is 9.994MHz/us and the ADC sample rate is 6250kHz. So the radar's maximum sensing range can reach $\frac{3 \times 10^8m/s \times 6250kHz}{9.994MHz/us \times 2}=93.8m$. The angle of the radar's field of view (FoV) is about $120^{\circ}$, which is large enough to cover the experimental scene. The ADC samples from the radar are captured by a Ti DCA1000EVM data acquisition board \cite{DCA} and then transmitted to a computer with an Intel Core i9-11900H 2.5GHz CPU for processing.

The experiment scenario is shown in Fig. \ref{fig:experiment}. The radar is fixed horizontally on a tripod mount and is calibrated with a corner reflector in advance. A six-wing UAV with three blades per propeller is used as the detection target. The UAV weights about 8kg and each blade is about 25cm long. It has a maximum velocity of 4m/s. The UAV is equipped with a \reviewII{Real-time Kinematic Positioning (RTK)} module \cite{rtk} to provide the ground truth of the UAV's location, which has a cm-level precision. 
We collect the reflected signals of the UAV from different flight trajectories, at different altitudes and different velocities. These reflected signals are first processed to obtain the tracking results and the corresponding Doppler-Time diagrams. The tracking results are aligned with the ground truth by their timestamps and trajectory features. Then the Doppler-Time diagrams are split into segments of 3.6s duration. We collect more than 4000 seconds of signals under different settings and generate over 1000 segments. We further generate the same amount of the other objects' segments by recording the Doppler-Time diagrams of kites, birds, balloons and shaking trees. \reviewII{We place the radar under balloons, kites, and shaking trees for data collection, and near a nest for bird data collection}. These two types of segments together form our dataset. In our implementation, we use the random 70\% of the segments as the training set and the rest 30\% as the test set. The model is trained using the Adam optimizer with a learning rate of 0.00005 and a batch size of 10.

\subsection{Methodology}

\review{We use \textit{accuracy}, \textit{precision}, \textit{recall} and F1-\textit{score} as the performance metrics to evaluate the performance of \systemname{}. The accuracy, precision, recall and F1-score are calculated from True Positive (TP), True Negative (TN), False Positive (FP) and False Negative (FN). They are calculated as follows: $accuracy = \frac{TP + TN}{TP + FP + FN + TN}$, $precision = \frac{TP}{TP + FP}$, $recall = \frac{TP}{TP + FN}$ and $F1$-$score = \frac{2*precision*recall}{precision + recall}$.}


To clearly show the whole detection process, we also evaluate the UAV tracking performance. We measure the tracking accuracy with the average relative range error. It calculates as follows:
\begin{equation}
\begin{aligned}
    \frac{1}{N} \sum_{n=1}^{N} \frac{|G(n) - T(n)|}{G(n)}
    \end{aligned}
\end{equation}
where $G(n)$ and $T(n)$ represent the actual range and the tracking range at the $n$-th sampling timestamp, respectively. $N$ is the number of the sampling timestamps in a trace.

\subsection{Overall Performance}

\begin{figure*}[ht]
  \centering 
\begin{minipage}{0.22\textwidth}
\centering 
\includegraphics[width=\textwidth]{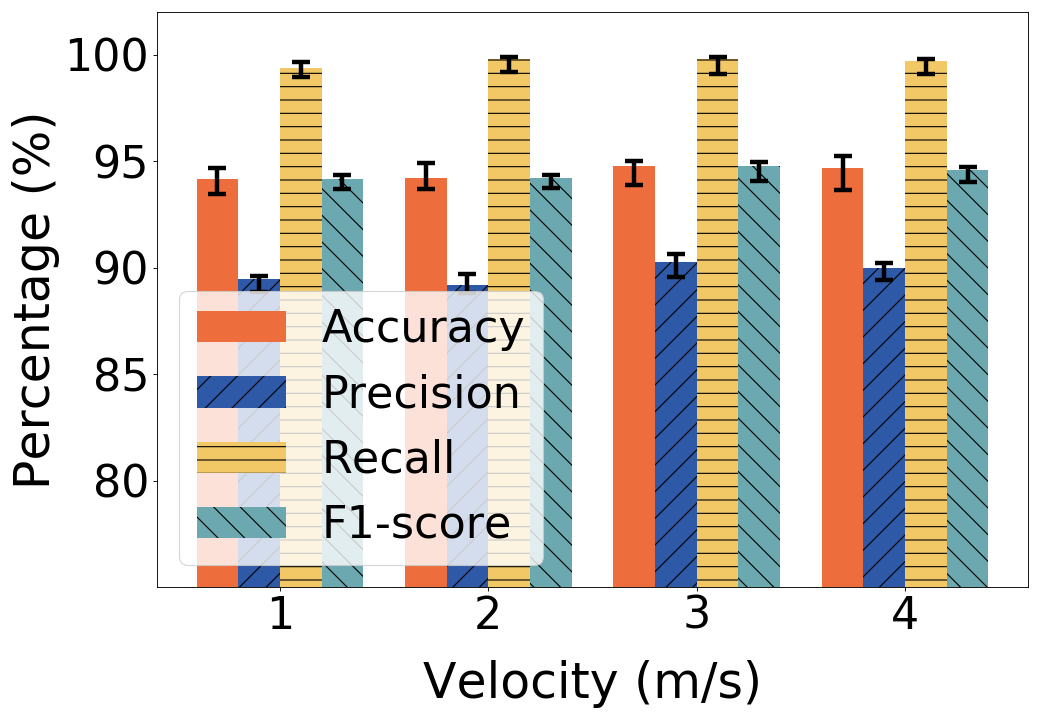}
\vspace{-0.5cm}
\caption{\review{The impact of UAV velocity on detection results}}
  \label{fig:velocity_detection_error}
\end{minipage}
\hspace{0.2cm}
\begin{minipage}{0.22\textwidth}
\centering 
\includegraphics[width=\textwidth]{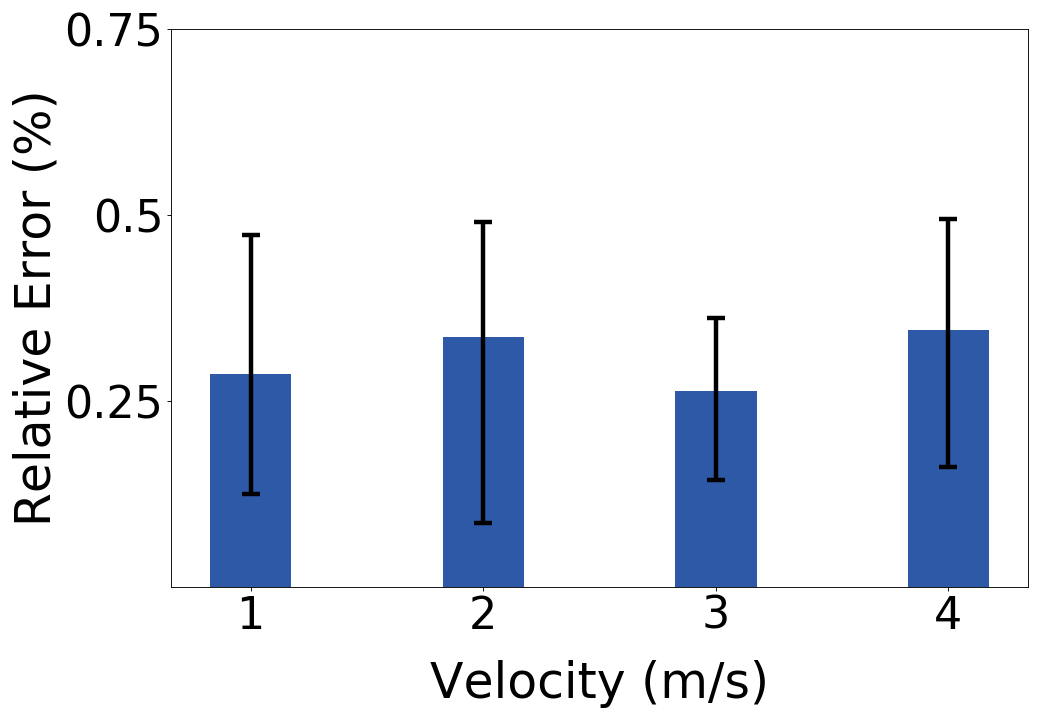}
\vspace{-0.5cm}
\caption{The impact of UAV velocity on tracking results}
  \label{fig:velocity_range_error}
\end{minipage}
\hspace{0.2cm}
 \begin{minipage}{0.22\textwidth}
 \centering
 \includegraphics[width=\textwidth]{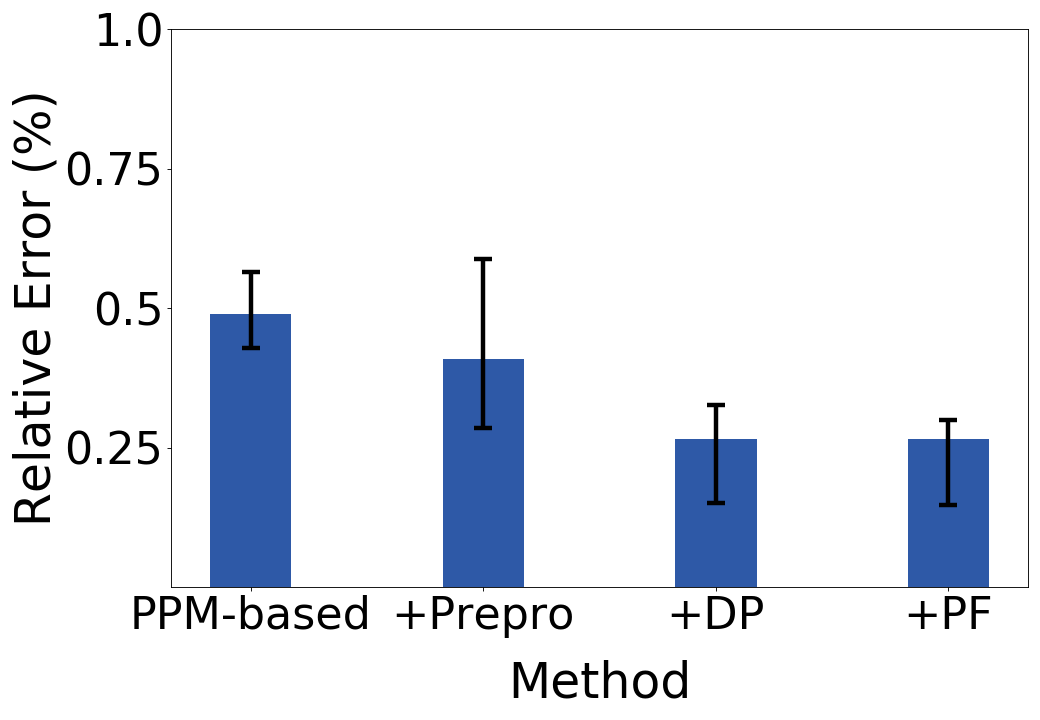}
 \vspace{-0.5cm}
  \caption{Ablation study on UAV tracking}
  \label{fig:ablation_study_range}
\end{minipage}
\hspace{0.2cm}
\begin{minipage}{0.22\textwidth}
\centering 
\includegraphics[width=\textwidth]{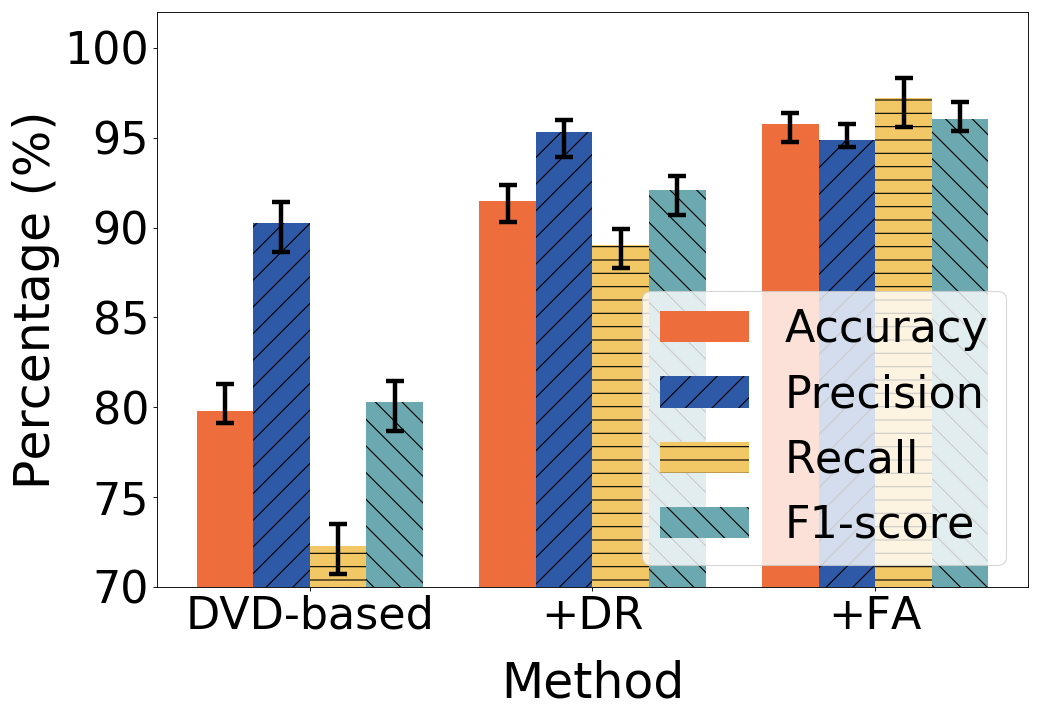}
\vspace{-0.5cm}
\caption{\review{Ablation study on UAV identification}}
  \label{fig:ablation_study_detection}
\end{minipage}
\vspace{-0.5cm}
\end{figure*}

\subsubsection{Detection accuracy}
We first evaluate the overall detection performance of \systemname{}. The UAV is controlled to perform different flight trajectories. The flight altitude varies from 10m to 80m. The flight trajectories at each altitude include hovering, horizontal flight, vertical flight and flying with a horizontal square. \reviewII{The radar is placed on the ground and vertically aligned with the horizontal center of the trajectories.} The detection results at different altitudes are shown in Fig. \ref{fig:overall_detection}. \review{The overall detection accuracy of \systemname{} is 95.8\% (corresponding with 92.6\% of precision, 97.2\% of recall and 94.8\% of F1-score). With the UAV altitude increasing, the detection accuracy decreases from 98.3\% to 91.3\%. When the UAV altitude is 10m, \systemname{} achieves an accuracy up to 98.3\%, a precision of 99.8\%, a recall of 96.8\% and an F1-score of 98.3\%. When the altitude increases to 80m, the detection performance falls to an accuracy of 91.3\%, a precision of 84.8\%, a recall of 94.8\% and an F1-score of 89.5\%.} As the UAV altitude increases, the PMM feature extracted from the received signal becomes weaker and sometimes incomplete, which leads to degradation in the detection performance. \systemname{} keeps a high recall, which means that it can effectively detect most of the UAVs even when the UAVs are at height.



\subsubsection{Tracking accuracy}
We further evaluate the tracking performance. We compare \systemname{} with the intensity-based tracking approach commonly used in previous works such as WaveEar \cite{waveear} and mmTrack \cite{mmtrack}, which select the largest value of the Range-FFT result as the tracking result. For consistency and fairness, we also perform preprocessing, dynamic programming, and particle filtering on the result of the compared approach. The tracking results of \systemname{} at different altitudes are shown in Fig. \ref{fig:overall_range_error}. \review{The results show that mmHawkeye can achieve a detection range of 80m, which is much higher than the detection range of 30m with the intensity-based approach.} When the UAV altitude is below 30m, our tracking algorithm can achieve a more accurate range estimation compared with the intensity-based approach. The average relative range error of \systemname{} decreases as the altitude increases. The reason is that the average range error of \systemname{} remains less than 10cm within 30m, resulting in a reduction in average relative tracing error. When the UAV altitude is above 30m, the intensity-based approach fails completely due to the low SNR of the reflected signal. 
In comparison, \systemname{} keeps accurate and has less than 2\% relative range error within 80m. This is mainly attributed to the utilization the UAV's PMM feature, which helps to efficiently distinguish UAVs from the environment noise.



\subsection{The Impact of Different Factors}

\subsubsection{The impact of the UAV's trajectory}
In this experiment, We evaluate the detection performance under different trajectories of the UAV. The results are shown in Fig. \ref{fig:trajectory_detect_error}. The detection accuracy varies slightly from 95.3\% to 96.6\% across different types of  trajectories. Since the micro-motion of the UAV's propellers always exists in different trajectories, the PMM feature remains consistent and can be used to resist the impact of the unpredictable UAV's motion and in turn keeps the good detection performance.

We also evaluate the tracking performance under different trajectories of the UAV. The average relative range errors under different trajectories at three altitudes (20m, 40m, 60m) are shown in Fig. \ref{fig:trajectory_range_error}. The range errors vary slightly with the types of UAV trajectory at each altitude and the range errors with the vertical flight are the lowest. The reason is that when the UAV is flying vertically, its radial velocity is the largest and the calculated folding result is more distinguishable. Since \systemname{} utilizes the PMM feature of the UAV to track it, we can resist the impact of the UAV's motion and keeps a reliable tracking result. 

\subsubsection{The impact of the UAV's velocity}
In this experiment, We evaluate the impact of the UAV's velocity on the detection performance. Due to the page limit, we only show the detection results of different UAV velocities at the altitude of 40m in Fig. \ref{fig:velocity_detection_error}. The detection accuracy varies slightly from 94.1\% to 94.8\% with the different UAV velocities. Since \systemname{} removes velocity-related features from the obtained Doppler-Time diagram, the velocity of the UAV has little impact on the detection performance.

We further evaluate the impact of the UAV's velocity on the tracking performance. The average relative range errors at different  velocities are shown in Fig. \ref{fig:velocity_range_error}. The relative range errors vary from 0.26\% to 0.35\% at different UAV velocities.
This result demonstrates that our UAV tracking algorithm remains effective applied under different UAV velocities. The reason is that our algorithm focuses on the frequency peaks caused by the UAV's PMM rather than the body velocity, which makes the tracking result stable even when the UAV's velocity is varied.

\subsection{Ablation Study}

\review{As the feature extraction module in mmHawkeye serves the other two modules and cannot be evaluated separately, we conduct ablation study on UAV tracking and UAV identification modules separately. The impact of feature extraction module can be verified by comparing with the experimental results of intensity-based approach.}

\subsubsection{UAV tracking}
This section evaluates the performance of the UAV tracking algorithm. We select the reflected signals when the UAV altitude is 40m as the target of processing. Then we respectively adopt the tracking method (1) based on the maximum value of the folding results (PMM-based), (2) based on the maximum value of the preprocessed folding results (+Prepro), (3) with dynamic programming of the preprocessed folding results (+DP), (4) with dynamic programming and particle filtering of the preprocessed folding results (+PF). The method +PF is the algorithm used in \systemname{}. 

The experiment result is shown in Fig. \ref{fig:ablation_study_range}. The results demonstrate that both the preprocessing and the dynamic programming significantly reduce the range error \review{by 16.4\% and 35.12\%, respectively}. The reason is that the preprocessing removes the static background noise from the folding results, and the dynamic programming considers the trajectory continuity and can resist the impact of unpredictable UAV motions. The results also demonstrate that the particle filtering can effectively reduce the variation of the range error \review{by 13.83\%}, since it can effectively resist the impact of the unpredictable motions and the local dynamic noise. \review{Furthermore, the intensity-based approach fails when the UAV altitude is 40m, and our simple PMM-based method can achieve a relative range error of 0.49\%. This means that spectrum folding can significantly improve UAV tracking accuracy.}

\subsubsection{UAV identification}
This subsection evaluates the performance of the UAV identification algorithm. For the obtained Doppler-Time diagrams, we respectively perform (1) no processing (DTD-based), (2) DC removal (+DR), (3) DC removal and feature alignment (+FA) to obtain the corresponding results. The results of each method are used to train the corresponding detection network and test the detection performance. The method +FA is the algorithm used in \systemname{}.

The detection results of these methods are shown in Fig. \ref{fig:ablation_study_detection}. The results demonstrate that the DC removal can significantly improve the detection accuracy \review{by 14.62\%}, since it can remove a large portion of the DC noise and make the PMM features easier to be learned. The results also show that the feature alignment can effectively improve the detection accuracy \review{by 4.68\%}, since it can remove the velocity-related features from the Doppler-Time diagrams and make the periodicity of the PMM feature easier to be identified.

\section{Discussion}


\subsection{Short-range Detection}
When the distance between the UAV and the radar is close (e.g. below 10m), the UAV can no longer be regarded as a whole and each part may occupy a different range bin. In this case, there will be multiple PMM features in different range bins in the Range-Doppler spectrum. We can utilize these features to further track and sense the UAV. Note that probably the UAV has already been detected by \systemname{} before it arrives in such a close range.

\vspace{-0.1cm}
\review{
\subsection{Practical Deployment}
In practical scenarios, UAVs can intrude into personal space at height or from around. However, considering that low-altitude UAVs are easy to be detected manually or by other approaches, and these UAVs have limited viewing angles, there are very few cases of actual intrusion into personal space from low altitude. In this case, we usually deploy the radar horizontally to detect the presence of UAVs at height rather than the presence of UAVs from around. 
}



\section{Related Work}

\review{
\subsection{UAV Detection with mmWave Radar}
With the development of COTS mmWave radar and the growing concern about UAVs, there have been some works utilizing mmWave radar to track and identify small UAVs. \cite{drone} captures the UAV 3-D motion with a novel deep neural network and achieves decimeter-level tracking accuracy within 5m. \cite{2022activity} utilizes the reflected mmWave signal intensity to calculate the distance and the elevation angle between UAV and radar, realizing the UAV tracking within 10m. It further extracts the micro-Doppler signatures from the captured UAV and achieves activity classification accuracy of 95\%. \cite{2021detection} employs the constant false alarm rate (CFAR) detector on the Range-Doppler spectrum and achieves a maximum detection range of about 40m. However, its performance decreases significantly when the UAV is hovering as the Doppler feature is less significant. In conclusion, none of them can achieve long-range UAV detection with the COTS mmWave radar due to the limited reflected signal intensity. To solve this problem, we directly utilize the PMM feature of the flying UAV rather than the signal intensity to achieve long-range UAV detection.}

\vspace{-0.1cm}
\subsection{UAV Detection with Other Devices}
There have been proposals to utilize sound \cite{acoustic_droneprint}, visual information \cite{video_vision}, and RF signals \cite{RF_matthan} to achieve passive non-cooperative UAV detection. For example, DronePrint \cite{acoustic_droneprint} proposes to detect a drone according to its acoustic signatures. Such approaches are susceptible to the environment noise and have a limited sensing range. \cite{video_vision} uses cascades of boosted classifiers on the collected videos to detect the UAV and achieve distance estimation. Such video-based approaches are easily affected by illumination conditions and complex backgrounds. Matthan \cite{RF_matthan} detects the presence of UAVs by monitoring the unique characteristics of the received WiFi signal. It achieves over 80\% detection accuracy within 600m range. However, \review{it requires non-cooperative UAVs to actively send WiFi signals, which renders silent UAVs undetectable.}


\section{Conclusion}
Defending against uninvited UAVs is an increasingly important problem nowadays. This paper presents our study on passive UAV detection. Our proposal named \systemname{} is a mmWave-sensing based approach that has broad applicability and satisfactory accuracy. \systemname{} particularly tackles the problem induced by low SNR signals and achieves long-range detection. Extensive experiments with the implemented prototype demonstrate that \systemname{} is accurate and reliable under various settings.

\section*{Acknowledgment}
We thank our anonymous shepherd and reviewers for their insightful comments. This work is supported by the National Science Fund of China under grant No. U21B2007 and Tsinghua University - Meituan Joint Institute for Digital Life.

\bibliographystyle{IEEEtran}
\bibliography{ref.bib}

\begin{thebibliography}{10}
\providecommand{\url}[1]{#1}
\csname url@samestyle\endcsname
\providecommand{\newblock}{\relax}
\providecommand{\bibinfo}[2]{#2}
\providecommand{\BIBentrySTDinterwordspacing}{\spaceskip=0pt\relax}
\providecommand{\BIBentryALTinterwordstretchfactor}{4}
\providecommand{\BIBentryALTinterwordspacing}{\spaceskip=\fontdimen2\font plus
\BIBentryALTinterwordstretchfactor\fontdimen3\font minus
  \fontdimen4\font\relax}
\providecommand{\BIBforeignlanguage}[2]{{%
\expandafter\ifx\csname l@#1\endcsname\relax
\typeout{** WARNING: IEEEtran.bst: No hyphenation pattern has been}%
\typeout{** loaded for the language `#1'. Using the pattern for}%
\typeout{** the default language instead.}%
\else
\language=\csname l@#1\endcsname
\fi
#2}}
\providecommand{\BIBdecl}{\relax}
\BIBdecl

\bibitem{Harry}
P.~Six, ``Prince harry and meghan markle call cops over drones flying over
  home,'' 2020,
  \url{https://pagesix.com/2020/05/28/prince-harry-and-meghan-markle-call-cops-after-drones-fly-over-home/}.

\bibitem{drug}
CNA, ``4 arrested after drone carrying drugs spotted over kranji reservoir
  park,'' 2020,
  \url{https://www.channelnewsasia.com/singapore/drone-drug-trafficking-arrest-kranji-reservoir-park-655706}.

\bibitem{liu2010long}
Y.~Liu \emph{et~al.}, ``Long-term large-scale sensing in the forest: recent
  advances and future directions of greenorbs,'' \emph{Frontiers of Computer
  Science in China}, 2010.

\bibitem{Australian}
BBC, ``Australian triathlete injured after drone crash,'' 2020,
  \url{https://www.bbc.com/news/technology-26921504}.

\bibitem{acoustic_droneprint}
H.~Kolamunna \emph{et~al.}, ``Droneprint: acoustic signatures for open-set
  drone detection and identification with online data,'' in \emph{ACM UbiComp},
  2021.

\bibitem{he2023acoustic}
Y.~He \emph{et~al.}, ``Acoustic localization system for precise drone
  landing,'' \emph{IEEE TMC}, 2023.

\bibitem{yimiao2023aim}
Y.~Sun \emph{et~al.}, ``Aim: Acoustic inertial measurement for indoor drone
  localization and tracking,'' in \emph{ACM SenSys}, 2022.

\bibitem{weiguo2023micnest}
W.~Wang \emph{et~al.}, ``Micnest: Long-range instant acoustic localization of
  drones in precise landing,'' in \emph{ACM SenSys}, 2022.

\bibitem{video_vision}
F.~G{\"o}k{\c{c}}e \emph{et~al.}, ``Vision-based detection and distance
  estimation of micro unmanned aerial vehicles,'' \emph{Sensors}, 2015.

\bibitem{RF_matthan}
P.~Nguyen \emph{et~al.}, ``Matthan: Drone presence detection by identifying
  physical signatures in the drone's rf communication,'' in \emph{ACM MobiSys},
  2017.

\bibitem{omnitrack}
C.~Jiang \emph{et~al.}, ``3d-omnitrack: 3d tracking with cots rfid systems,''
  in \emph{ACM/IEEE IPSN}, 2019.

\bibitem{ram}
M.~A. A.~H. Khan \emph{et~al.}, ``Ram: Radar-based activity monitor,'' in
  \emph{IEEE INFOCOM}, 2016.

\bibitem{radar_lband}
M.~Jahangir and C.~Baker, ``Persistence surveillance of difficult to detect
  micro-drones with l-band 3-d holographic radar,'' in \emph{IEEE RADAR}, 2016.

\bibitem{rfwash}
A.~Khamis \emph{et~al.}, ``Rfwash: a weakly supervised tracking of hand hygiene
  technique,'' in \emph{ACM SenSys}, 2020.

\bibitem{mmface}
W.~Xu \emph{et~al.}, ``Mask does not matter: Anti-spoofing face authentication
  using mmwave without on-site registration,'' in \emph{ACM MobiCom}, 2022.

\bibitem{ambiear}
J.~Zhang \emph{et~al.}, ``Ambiear: mmwave based voice recognition in nlos
  scenarios,'' \emph{ACM UbiComp}, 2022.

\bibitem{rf-scg}
U.~Ha \emph{et~al.}, ``Contactless seismocardiography via deep learning
  radars,'' in \emph{ACM MobiCom}, 2020.

\bibitem{drone}
P.~Zhao \emph{et~al.}, ``3d motion capture of an unmodified drone with
  single-chip millimeter wave radar,'' in \emph{IEEE ICRA}, 2021.

\bibitem{jin2022passive}
M.~Jin \emph{et~al.}, ``A passive eye-in-hand" camera" for miniature robots,''
  in \emph{ACM SenSys}, 2022.

\bibitem{2021detection}
P.~J.~B. Morris and K.~Hari, ``Detection and localization of unmanned aircraft
  systems using millimeter-wave automotive radar sensors,'' \emph{IEEE Sensors
  Letters}, 2021.

\bibitem{jin2023}
M.~Jin \emph{et~al.}, ``Fast, fine-grained, and robust grouping of rfids,'' in
  \emph{ACM MobiCom}, 2023.

\bibitem{2022activity}
N.~R. Beeram \emph{et~al.}, ``Activity classification of an unmanned aerial
  vehicle using tsetlin machine,'' in \emph{IEEE ISTM}, 2022.

\bibitem{he2019red}
Y.~He \emph{et~al.}, ``Red: Rfid-based eccentricity detection for high-speed
  rotating machinery,'' \emph{IEEE TMC}, 2019.

\bibitem{yang2022wiimg}
K.~Yang \emph{et~al.}, ``Wiimg: Pushing the limit of wifi sensing with low
  transmission rates,'' in \emph{IEEE SECON}, 2022.

\bibitem{mmVib}
C.~Jiang \emph{et~al.}, ``mmvib: micrometer-level vibration measurement with
  mmwave radar,'' in \emph{ACM MobiCom}, 2020.

\bibitem{analysis}
K.-B. Kang \emph{et~al.}, ``Analysis of micro-doppler signatures of small uavs
  based on doppler spectrum,'' \emph{IEEE TAES}, 2021.

\bibitem{6843}
T.~I. Incorporated, ``Hardware setup for mmwaveicboost and iwr6843iskods,''
  \url{https://training.ti.com/hardware-setup-mmwaveicboost-and-antenna-module?keyMatch=IWR6843ISKODS},
  2022.

\bibitem{freebee}
S.~M. Kim and T.~He, ``Freebee: Cross-technology communication via free
  side-channel,'' in \emph{ACM MobiCom}, 2015.

\bibitem{guo2020wizig}
X.~Guo \emph{et~al.}, ``Wizig: Cross-technology energy communication over a
  noisy channel,'' \emph{IEEE/ACM TON}, 2020.

\bibitem{advanced}
S.~V. Vaseghi, \emph{Advanced digital signal processing and noise
  reduction}.\hskip 1em plus 0.5em minus 0.4em\relax John Wiley \& Sons, 2008.

\bibitem{particle}
G.~Bielsa \emph{et~al.}, ``Indoor localization using commercial off-the-shelf
  60 ghz access points,'' in \emph{IEEE INFOCOM}, 2018.

\bibitem{lstm}
S.~Hochreiter and J.~Schmidhuber, ``Long short-term memory,'' \emph{Neural
  computation}, 1997.

\bibitem{DCA}
T.~I. Incorporated, ``Real-time data-capture adapter for radar sensing
  evaluation module,'' \url{http://www.ti.com/tool/DCA1000EVM}, 2020.

\bibitem{rtk}
Wikipedia, ``Real-time kinematic positioning,''
  \url{https://en.wikipedia.org/wiki/Real-time_kinematic_positioning}, 2022.

\bibitem{waveear}
C.~Xu \emph{et~al.}, ``Waveear: Exploring a mmwave-based noise-resistant speech
  sensing for voice-user interface,'' in \emph{ACM MobiSys}, 2019.

\bibitem{mmtrack}
C.~Wu \emph{et~al.}, ``mmtrack: Passive multi-person localization using
  commodity millimeter wave radio,'' in \emph{IEEE INFOCOM}, 2020.

\end{thebibliography}

\end{document}